# Lagrange Multipliers

# and

# Third Order

# Scalar-Tensor Field Theories

by

Gregory W. Horndeski

2814 Calle Dulcinea
Santa Fe, NM 87505-6425
e-mail:
horndeskimath@gmail.com

September 10, 2016

**ABSTRACT**

In a space of 4-dimensions, I will examine constrained variational problems in which the Lagrangian, and constraint scalar density, are concomitants of a (pseudo-Riemannian) metric tensor and its first two derivatives. The Lagrange multiplier for these constrained extremal problems will be a scalar field. For suitable choices of the Lagrangian, and constraint, we can obtain Euler-Lagrange equations which are second order in the scalar field and third order in the metric tensor. The effect of disformal transformations on the constraint Lagrangians, and their generalizations, is examined. This will yield other second order scalar-tensor Lagrangians which yield field equations which are at most of third order. No attempt is made to construct all possible third order scalar-tensor Euler-Lagrange equations in a 4-space, although nine classes of such field equations are presented. Two of these classes admit subclasses which yield conformally invariant field equations. A few remarks on scalar-tensor-connection theories are also presented.

**Section 1: Introduction**

About 15 years ago Professor Glen Schrank asked me to compute the variational derivative of $g^{1/2} R_{abcd}R^{abcd}$, in a space of 4-dimensions, subject to the constraint that the variation be through metrics for which the Gauss-Bonnet scalar density, GB, was zero, where

$$GB := g^{1/2}(R^2 + 4R_{ij}R^{ij} + R_{hijk}R^{hijk}) = \tfrac{1}{4}\, g^{1/2}\, \delta^{abcd}{}_{rstu} R^{rs}{}_{ab} R^{tu}{}_{cd}\,. \qquad \text{Eq.1.1}$$

(The notation used here is the same as that used in Horndeski & Lovelock[1], and Horndeski [2], with the exception of the symbol used for the generalized Kronecker delta.) I informed Schrank that such a variation could easily be performed through the introduction of a Lagrange multiplier, which in the present case, would be a scalar field, φ. The resulting constrained Lagrangian for the Schrank problem is

$$L_S := g^{1/2}\, R_{abcd}R^{abcd} + \varphi\, GB\,. \qquad \text{Eq.1.2}$$

The separate, unrestricted variations, of $g_{ij}$ and φ in $L_S$, will yield the EL (:= Euler-Lagrange) equations equivalent to those obtained from a restricted variation of $g_{ij}$ in $g^{1/2} R_{abcd}R^{abcd}$, subject to the constraint that GB = 0. A lengthy calculation, using the formalism presented in [1], noting that $E^{ab}(GB) = 0$, shows that

$$E^{ab}(L_S) = -2(g^{1/2})[(R^{acdb} + R^{bcda})_{|cd} - R_k{}^a{}_{mh} R^{kbmh} + \tfrac{1}{4} g^{ab} R^{rstu} R_{rstu}] +$$

$$+ 2(g^{1/2})[g^{ab} \Box \varphi \, R - 2g^{ab} \varphi_{pq} R^{pq} - \varphi^{ab} R + 2\varphi^{pa} R^b{}_p + 2\varphi^{pb} R^a{}_p - 2\Box\varphi \, R^{ab} +$$

$$+ 2\varphi_{pq} R^{apbq}] , \qquad \text{Eq.1.3}$$

and

$$E(L_S) = -(g^{1/2})[R^2 - 4R_{ij} R^{ij} + R_{hijk} R^{hijk}] , \qquad \text{Eq.1.4}$$

where, for simplicity, I will not use a vertical bar to denote the covariant derivative of $\varphi$. So, e.g., $\varphi^{pa} := \varphi^{|pa}$, and $\Box\varphi := \varphi^a{}_a$. Evidently, these equations are fourth order in $g_{ab}$ and second order in $\varphi$.

Now recently there has been some interest in third order scalar-tensor field equations which would serve to generalize the results presented in [2]. Some results dealing with this subject can be found in Zumalacárregui and Garcia-Bellido [3], Deffayet, *et al.*, [4], Gleyzes, *et al.*, [5], [7], Lin, *et al.*, [6] and Crisostomi, *et al.*, [8]. We see that the equations given by $E^{ab}(L_S) = 0$, and $E(L_S) = 0$, would be higher order scalar-tensor field equations, but they are of fourth order in $g_{ij}$. Is it possible to arrive at third order scalar-tensor field equations through the use of a constrained metric variational approach?

To see that it is, let us consider a theorem of Lovelock's. In [9] Lovelock has demonstrated that in 4-dimensional space the most general Lagrangian scalar density of the form

$$L = L(g_{ab}; g_{ab,c}; g_{ab,cd}) \qquad \text{Eq.1.5}$$

which is such that $E^{ab}(L)$ is at most of second order in $g_{ab}$ is

$$L_L := \alpha_1 g^{1/2} + \alpha_2 g^{1/2} R + \alpha_3 GB + \alpha_4 P, \qquad \text{Eq.1.6}$$

where $\alpha_1$, $\alpha_2$, $\alpha_3$ and $\alpha_4$ are real constants, and P is the (4-dimensional) Pontrjagin Lagrangian[10]

$$P := \varepsilon^{hijk} R^{lm}{}_{hi} R_{lmjk}. \qquad \text{Eq.1.7}$$

(When considering P, I shall assume that we are working on an orientable manifold with charts which are such that on the overlap of their domains the Jacobian is positive. Thus $\varepsilon^{abcd}$ will be a contravariant tensor density.) Since GB and P yield identically vanishing EL tensors we see, as Lovelock has proved, that in a space of 4-dimensions, the Einstein field equations with cosmological constant are the most general second order field equations derivable from a Lagrangian scalar density of the form given in Eq.1.5.

The reason that GB and P yield identically vanishing EL tensors is because both are locally expressible as divergences, provided one considers vector densities that are not only concomitants of $g_{ab}$, but also include other fields. GB can locally be expressed as a divergence using a vector density built from $g_{ab}$ and a vector field, along with their derivatives. P can locally be expressed as a divergence using a vector density built from $g_{ab}$ and a local parallelization, along with their derivatives.

Classical tensorial proofs of these facts can be found in Horndeski [11] and [12]. For a more contemporary, index-free approach to GB and P, along with their topological significance, please see Kobayashi and Nomizu [13].

Upon examining Eq.1.6 we note that if $\alpha_1$, $\alpha_2$, $\alpha_3$ and $\alpha_4$ were functions of $\varphi$, then the $\alpha_4$ P Lagrangian would be the only one of the four not found among the Lagrangians presented in [1] and [2], that yield second order scalar-tensor EL equations, even though $E(\alpha_4 P)$ is zeroth order in $\varphi$ and second order in $g_{ab}$. This suggests that it might be of interest to examine constrained metric variational problems with a Lagrangian of the form, $L + \varphi P$, where L is of the type presented in Eq.1.5. I will do this in the next section, where it will be shown that $\varphi$ P generates EL equations of third order in $g_{ab}$ and second order in $\varphi$. In section 3, I shall examine the effect of the simplest disformal transformation on $\varphi P$, and we shall see that we can generate other Lagrangians which yield EL equations of third order in the derivatives of $g_{ab}$ and $\varphi$. The penultimate section will deal with scalar-tensor-connection theories generated by the Lagrangians presented in sections 2 and 3. The purpose of introducing the affine connection will be to reduce the differential order of the EL equations, and thereby hopefully avoid ghosts, which can plague $3^{rd}$ order theories, due to the work of Ostrogradski [14]. The fifth, and final section of this paper, begins by summarizing the highlights of what was done in the previous sections. I then go on to show how these results can be used to construct even more second order scalar-tensor Lagrangians which yield third order EL equations. This then leads to a discussion of how one can proceed to determine all possible third order scalar-tensor field equations which are derivable from a scalar-tensor Lagrangian in a 4-space.

Before we embark upon our study of φ P and its cohorts, we need to take another look at GB. I mentioned above that when $\alpha_3$ is an arbitrary function of φ, then the Lagrangian $\alpha_3(\varphi)$ GB yields second order EL equations. However, if U is an arbitrary differentiable function of φ and ρ, where $\rho := g^{ab} \varphi_a \varphi_b$, then U GB is not found amongst the Lagrangians presented in [1] and [2], that yield second order EL equations. That is because both $E^{ab}$(U GB) and E(U GB) are third order. In fact using the formalism presented in [1] it can be shown that

$$E^{ab}(U\ GB)\ =\ g^{1/2}\ U_{|hk}\ \delta^{akpq}{}_{vwtu}\ g^{bv}\ g^{hw}\ R^{tu}{}_{pq} + U_\rho\ \varphi^a\ \varphi^b\ GB$$

$$= 2\ g^{1/2}\ [g^{ab}\ \Box U\ R - 2g^{ab}\ U_{|pq}\ R^{pq} - U^{|ab}\ R + 2U^{|pa}\ R^b{}_p + 2U^{|pb}\ R^a{}_p +$$

$$- 2\ \Box U\ R^{ab} + 2U_{|pq}\ R^{apbq}\ ] + U_\rho\ \varphi^a\ \varphi^b\ GB\ ,\quad\quad\quad Eq.1.8$$

and

$$E(U\ GB)\ = [2U_\rho\ \varphi^a\ GB]_{|a} - U_\varphi\ GB\ ,\quad\quad\quad Eq.1.9$$

where partial derivatives of U with respect to φ and ρ are denoted by the subscripts φ and ρ. Thus we see that $E^{ab}$(U GB) will be third order in φ if $U_\rho \neq 0$, and E(U GB) will be third order in $g_{ab}$ if $U_\rho \neq 0$. Throughout the remainder of this paper I shall be remarking about the properties of U GB and how they compare with the Lagrangians generated by P.

I must mention that in [15] Bettoni and Zumalacárregui, discuss the possibility of using U GB to generate so-called, Beyond Horndeski Theories. But they do not do much with that Lagrangian. However, in [16] Ezquiaga, *et al.*, introduce a differential form approach to study third order scalar-tensor field theories. Employing this ingenious approach, they study U GB, along with many of the other Lagrangians that I will present in this paper. My work here was developed independently of what they recently accomplished in [16], and should provide the reader with two different approaches to some of the same material.

In passing I would like to point out that what I am doing here differs significantly from what Sáez-Gómez does in [17]. There he considers Lagrangians which are at most of second order in $g_{ab}$ and $\varphi$, and then introduces a second scalar field as a Lagrange mulitplier, to perform a constrained variation of his original scalar-tensor Lagrangian.

**Section 2: Constrained Metric Variations with P = 0, and More**

If we want to vary a Lagrangian of the form presented in Eq.1.5, subject to the constraint that P = 0, then we need to investigate the Lagrangian

$$L = L + \varphi P \qquad \text{Eq.2.1}$$

where the scalar field φ acts as a Lagrange undetermined multiplier. *E.g.,* we could choose L in Eq.2.1 to be a Lagrangian of the form $g^{1/2}$ f(R) ( which was first studied by Buchdahl [18]), and thereby obtain a scalar-tensor f(R) theory. (For a comprehensive survey of the study of $g^{1/2}$ f(R) Lagrangians, and their relationship to scalar-tensor theories, *see* Sotiriou and Faraoni [19].) We will not be concerned with the actual form of L in Eq.2.1. Our attention will be concentrated on φ P, which is a special case of the Lagrangian

$$L_P := J P, \qquad \text{Eq.2.2}$$

where J is any differentiable function of φ and ρ.

Using the formalism presented in [2] we find, noting that $E^{ab}(P) = 0$, that

$$E^{ab}(L_P) = -2 J_{|h} (\partial P / \partial g_{ab,hk})_{|k} - J_{|hk} (\partial P / \partial g_{ab,hk}) + J_\rho \varphi^a \varphi^b P \qquad \text{Eq.2.3}$$

and

$$E(L_P) = 2 J_{\varphi\rho} \rho P + 4 J_{\rho\rho} \varphi^h \varphi^k \varphi_{hk} P + 2 J_\rho \Box\varphi P + 2 J_\rho \varphi^h P_{|h} - J_\varphi P . \qquad \text{Eq.2.4}$$

It is easily seen (noting Eq.3.50 in chapter 8 of Lovelock & Rund [20], and their conventions for carrying out derivatives with respect to $g_{ab}$, $g_{ab,c}$ and $g_{ab,cd}$, which will be used here) that

$$\partial P / \partial g_{ab,hk} = \varepsilon^{rsah} R_{rs}{}^{kb} + \varepsilon^{rsak} R_{rs}{}^{hb} + \varepsilon^{rsbh} R_{rs}{}^{ka} + \varepsilon^{rsbk} R_{rs}{}^{ha} . \qquad \text{Eq.2.5}$$

Using the second Bianchi identity, along with Eq.2.5, gives us

$$[\partial P / \partial g_{ab,hk}]_{|k} = -2\, \varepsilon^{rsah} R_{r\ |s}^{\ b} - 2\, \varepsilon^{rsbh} R_{r\ |s}^{\ a}\ . \qquad \text{Eq.2.6}$$

Combining Eqs. 2.3, 2.5 and 2.6, shows us that

$$E^{ab}(L_P) = 4[J_\varphi\, \varphi_h + 2\, J_\rho\, \varphi^l\, \varphi_{lh}][\varepsilon^{rsah} R_{r\ |s}^{\ b} + \varepsilon^{rsbh} R_{r\ |s}^{\ a}] +$$

$$-2[J_{\varphi\varphi}\, \varphi_h\, \varphi_k + 2\, J_{\varphi\rho}(\varphi^l \varphi_{lh}\, \varphi_k + \varphi^l \varphi_{lk}\, \varphi_h) + 4 J_{\rho\rho}\, \varphi^l\, \varphi_{lh}\, \varphi^m\, \varphi_{mk} + J_\varphi\, \varphi_{hk} + 2 J_\rho\, \varphi^l_{\ h}\, \varphi_{lk} +$$

$$+2\, J_\rho\, \varphi^l\, \varphi_{lhk}] \times [\varepsilon^{rsah} R_{rs}^{\ kb} + \varepsilon^{rsbh} R_{rs}^{\ ka}] + J_\rho\, \varphi^a\, \varphi^b\, P\ . \qquad \text{Eq.2.7}$$

Evidently, in general, $E^{ab}(L_P)$, is third order in $g_{ab}$ and $\varphi$, while $E(L)$ is 2$^{nd}$ order in $\varphi$ and third order in $g_{ab}$. We shall now examine two special cases of $E^{ab}(L_P)$ and $E(L_P)$; *viz.*, when $J = \varphi$, and $J = \rho$, giving $J = \varphi$ most of our attention.

Using Eqs. 2.4 and 2.7 we see that

$$E^{ab}(\varphi\, P) = 4\, \varphi_h\, [\varepsilon^{rsah} R_{r\ |s}^{\ b} + \varepsilon^{rsbh} R_{r\ |s}^{\ a}] - 2\, \varphi_{hk}\, [\varepsilon^{rsah} R_{rs}^{\ kb} + \varepsilon^{rsbh} R_{rs}^{\ ka}] \qquad \text{Eq.2.8}$$

and

$$E(\varphi\, P)\ = -P\ . \qquad \text{Eq.2.9}$$

$E^{ab}(\varphi\, P)$ of third order in $g_{ab}$, and second order in $\varphi$, which clearly explains why $\varphi\, P$ does not appear among the scalar-tensor Lagrangians presented in [2], which deals with second order EL equations. Another interesting property of $E^{ab}(\varphi\, P)$, is that it is trace-free. If one were now interested in performing a constrained variation of a Lagrangian, L, as given in Eq.1.5, subject to the constraint that $P = 0$, then the EL equations for that problem would be obtained by adding $E^{ab}(L)$ to $E^{ab}(\varphi\, P)$.

If L is any Lagrangian of the form

$$L = L(g_{ab}; g_{ab,c}; ...; \varphi; \varphi_{,a}; ...) \quad \text{Eq.2.10}$$

where the derivatives of $g_{ab}$ and $\varphi$ are of arbitrary (finite) order, then in Horndeski [21] and [22], I show that

$$E^{ab}(L)_{|b} = \tfrac{1}{2}\, \varphi^a\, E(L)\,. \quad \text{Eq.2.11}$$

Hence if the equation $E(L) = 0$ is satisfied, then $E^{ab}(L)$ is divergence free. This result is useful if one wants to equate $E^{ab}(L)$ to the energy-momentum tensor of matter, $T_M^{ab}$, since then the EL equations would imply that $T_M^{ab}{}_{|b} = 0$. Thus for $\varphi\, P$, we see that

$$E^{ab}(\varphi\, P)_{|b} = 0, \text{ if } E(\varphi P) = -P = 0\,.$$

The EL tensor $E^{ab}(\varphi\, P)$ should remind you of some famous third order, symmetric tensorial concomitant of the metric tensor, which is trace-free and divergence-free; *viz.,* the Cotton tensor [23]. This tensor only exists globally in an orientable 3-space, and locally is given by

$$C^{ij} = \varepsilon^{iab}\, R^{j}{}_{a|b} + \varepsilon^{jab}\, R^{i}{}_{a|b}\,. \quad \text{Eq.2.12}$$

$E^{ab}(\varphi\, P)$ seems to be a generalization of $C^{ij}$ into a 4-space. $C^{ij}$ has the property that a 3-space is conformally flat if and only if $C^{ij} = 0$ (see Cotton [23] or Eisenhart [24], for a proof of this). $C^{ij}$ is conformally invariant, and similarly, $E^{a}{}_{b}(\varphi P)$ is also conformally invariant, which a lengthy direct calculation demonstrates.

But there exists an easier way to prove that $E^{a}{}_{b}(\varphi\, P)$ is conformally invariant.

First one should note that

$$P = \varepsilon^{hijk} R^{lm}{}_{hi} R_{lmjk} = \varepsilon^{hijk} C^{lm}{}_{hi} C_{lmjk} ,  \qquad \text{Eq.2.13}$$

where $C_{hijk}$ is the Weyl tensor, which, in a 4-space, is given by

$$C_{hijk} = R_{hijk} + \tfrac{1}{2}( g_{hk} R_{ij} - g_{hj} R_{ik} + g_{ij} R_{hk} - g_{ik} R_{hj}) + \tfrac{1}{6} R (g_{hj} g_{ik} - g_{hk} g_{ij} ) .$$

Since $C^h{}_{ijk}$ is conformally invariant, it is clear from Eq.2.13 that P, and hence $\varphi$ P, are conformally invariant. A trivial calculation, for those who use the $\delta$ approach to computing variational derivatives, shows that if L is of the form given in Eq.2.10, and is invariant under a conformal transformation, then $E^a{}_b(L) = g_{bc} E^{ac}(L) = g_{bc}[\delta L/\delta g_{ac}]$, is conformally invariant. This can also be demonstrated by means of a direct, and more arduous calculation, using the definition of $E^{ab}(L)$ in terms of derivatives of L. Consequently, $E^a{}_b(\varphi P)$, and in fact $E^a{}_b( j(\varphi) P)$, are conformally invariant, where $j(\varphi)$ is an arbitrary differentiable function of $\varphi$.

We just saw that $\varphi P$ is conformally invariant, and $E^{ab}(\varphi P)$ is trace-free. This is no coincidence due to

**Proposition 2.1:** Let L be a Lagrange scalar density which is a concomitant of $g_{ab}$, $\varphi$, and their derivatives of arbitrary (finite) order, and assume that L is invariant under the conformal transformation $g_{ab} \to g'_{ab} := e^{2\sigma} g_{ab}$, where $\sigma$ is an arbitrary differentiable scalar field. So

$$L' := L(g'_{ab}; g'_{ab,c} ;...; \varphi; \varphi_{,a};...) = L(g_{ab}; g_{ab,c};...; \varphi; \varphi;...) . \qquad \text{Eq.2.14}$$

Under these assumptions $E^{ab}(L)$ is trace-free. Conversely, if $g_{ab} E^{ab}(L) = 0$, then $L'-L$ is a divergence, and hence $E^a{}_b(L)$ is conformally invariant.

**Proof:** Let $g(t)$ be a one parameter family of metric tensors with local components $g(t)_{ab} := (1-t)g_{ab} + tg'_{ab}$, $0 \leq t \leq 1$. So $g(t)$ is a convex combination of the metric tensors $g$ and $g'$. Using our expression for $g'_{ab}$ we see that $g(t)_{ab} = (1-t+te^{2\sigma})g_{ab}$. Thus $g(t)$ is a pseudo-Riemannian metric of the same signature as $g$, and is just a 1-parameter variation from $g$ to $g'$. We now define the 1-parameter family of Lagrangians $L(t)$ by

$$L(t) := L(g(t)_{ab}; g(t)_{ab,c}; ...; \varphi; \varphi_{,a}; ...) . \qquad \text{Eq.2.15}$$

Using an argument similar to the one used to prove Lemma 2.1 in Horndeski [25], we obtain the well known formula

$$\frac{d}{dt} L(t) = - E^{ab}(L(t)) \frac{dg(t)_{ab}}{dt} + \frac{d}{dx^a} V^a(t), \qquad \text{Eq.2.16}$$

where $E^{ab}(L(t))$ means the concomitant $E^{ab}(L)$, evaluated for the metric tensor $g(t)$. I shall not write out the expression for $V^a(t)$ in general, since that can be found in [25]. However, if $L$ is of second order in $g_{ab}$ the counterpart of Eq.2.16 would be

$$\frac{dL(t)}{dt} = -E^{ab}(L(t))\frac{dg(t)_{ab}}{dt} + \frac{d}{dx^c}\left\{\frac{\partial L}{\partial g_{ab,c}}(g(t)) \frac{dg(t)_{ab}}{dt} + \frac{\partial L}{\partial g_{ab,cd}}(g(t)) \frac{dg(t)_{ab,d}}{dt} +\right.$$

$$\left. - \frac{d}{dx^d}\left[\frac{\partial L}{\partial g_{ab,cd}}(g(t))\right]\frac{dg(t)_{ab}}{dt}\right\}. \qquad \text{Eq.2.17}$$

If L is conformally invariant it is obvious from Eq.2.15 that $\frac{dL(t)}{dt} = 0$.

So when $t = 0$, $\frac{dg(t)_{ab}}{dt} = (e^{2\sigma} - 1)g_{ab}$, and thus Eq.2.17 becomes

$$0 = -E^{ab}(L)(e^{2\sigma}-1)g_{ab} + \frac{d}{dx^c}\left\{\frac{\partial L}{\partial g_{ab,c}}(e^{2\sigma}-1)g_{ab} + \frac{\partial L}{\partial g_{ab,cd}}[2\sigma_{,d}e^{2\sigma}g_{ab} + (e^{2\sigma}-1)g_{ab,d}] \right.$$

$$\left. - \frac{d}{dx^d}\left[\frac{\partial L}{\partial g_{ab,cd}}\right](e^{2\sigma}-1)g_{ab}\right\} .$$   Eq.2.18

In this equation $\sigma_{,r}$ and $\sigma_{,rs}$ are essentially arbitrary. If we now let the $\frac{d}{dx^c}$ derivative act on the various terms in Eq.2.18, and then differentiate the resulting equation with respect to $\sigma_{,rs}$ and $\sigma_{,r}$ we can deduce that

$$0 = E^{ab}(L)(e^{2\sigma} - 1)g_{ab},$$

and hence $E^{ab}(L)$ is trace free if L is conformally invariant, and second order in $g_{ab}$. The proof of this result is similar when L is of arbitrary differential order in the metric tensor.

Now for the converse. If $E^{ab}(L)g_{ab} = 0$, then Eq.2.16 implies that

$$\frac{dL(t)}{dt} = \frac{d}{dx^a}V^a(t) .$$   Eq.2.19

Upon integrating Eq.2.19 with respect to t from 0 to 1 we obtain

$$L' - L = \frac{d}{dx^a}\int_0^1 V^a(t)dt ,$$

and so L' is equal to L, up to a divergence. ∎

The fact that if L is conformally invariant, then $E^{ij}(L)$ is trace-free, was proved by du Plesis for second order Lagrangians in [26]. His proof relies heavily upon L being a scalar density, while my proof does not require that. **Prop. 2.1** is also well-known to those who do Conformal Field Theory, see, *e.g.,* chapter 4 in Tong [27] for a proof in that context. Nevertheless, I wanted to present the above proof, since it is more in keeping with my approach to concomitant theory, which follows the Rund-Lovelock tradition presented in [20]. I would also like to mention that **Prop. 2.1** is valid in a space of arbitrary dimension, and remains valid if there are other fields in L, so long as they are not affected by the conformal transformation, which was the case with $\varphi$.

As an aside I would like to point out that even though the Cotton tensor is a symmetric, divergence-free concomitant of the metric tensor and its derivatives, it can not be obtained as the EL tensor from any (sufficiently differentiable) Lagrange scalar density which is only a concomitant of the metric and its derivatives. This fact was proved by Aldersley [28].

The above analysis has shown that $E^{ab}(\varphi P)$ and $E(\varphi P)$ are quite interesting EL tensors. We shall now proceed to examine a second special case of Lagrangians of the form $L_P = J \, P$. For this I choose $J = \rho$. Using Eqs.2.4 and 2.7 we see that

$$E^{ab}(\rho P) = 8\varphi^l \varphi_{lh}[\varepsilon^{rsah} R^b_{r\,|s} + \varepsilon^{rsbh} R^a_{r\,|s}] - 4[\varphi^l_h \varphi_{lk} + \varphi^l \varphi_{lhk}][\varepsilon^{rsah} R_{rs}{}^{kb} + \varepsilon^{rsbh} R_{rs}{}^{ka}] +$$
$$+ \varphi^a \varphi^b \, P \hspace{6cm} \text{Eq.2.20}$$

and

$$E(\rho P) = 2 \Box\varphi\, P + 2\, \varphi^h\, P_{|h} \quad . \qquad \text{Eq.2.21}$$

So $E^{ab}(\rho P)$ is of third order in $g_{ab}$ and $\varphi$, while $E(\rho P)$ is third order in $g_{ab}$ and second order in $\varphi$. Unlike $E^{ab}(\varphi P)$, $E^{ab}(\rho P)$ is not trace-free, and we have

$$g_{ab}\, E^{ab}(\rho P) = \rho P \ .$$

In fact $E^{ab}(J\, P)$ is trace-free, if and only if $J_\rho = 0$. Thus due to **Prop. 2.1**, $E^a{}_b(L_P)$ is conformally invariant if and only if $J_\rho = 0$, and thus $L_P = j(\varphi)\, P$ for an arbitrary differentiable function $j(\varphi)$. I will let

$$L_{PC} := j(\varphi)\, P, \qquad \text{Eq.2.22}$$

denote the subclass of Lagrangians of the $L_P$ type which are conformally invariant.

Although P is conformally invariant, GB is not. In fact a laborious calculation demonstrates that GB' is equal to GB plus a divergence in a 4-space. This result could be deduced immediately from **Prop.2.1**, since we know that $E^{ab}(GB) \equiv 0$, and hence $E^{ab}(GB)$ is trace-free. Consequently, **Prop.2.1** tells us that GB' and GB differ by a divergence. This is reassuring, since, in the case where the metric tensor is positive definite, we know that the integral of either one over a compact 4-manifold yields the same topological invariant.

Thus far we have two classes of Lagrangians, $L_{GB}$ and $L_P$ which yield third order EL equations. In the next section we shall examine the effects of disformal transformations on P and GB, and this will lead to more Lagrangians which yield third order EL equations.

**Section 3: The Effect of Disformal Transformations on P and GB**

In [29] Bekenstein introduced the notion of disformal transformations. My introduction to the subject was through a paper of Bettoni & Liberati [30]. Following them, for the purposes of scalar-tensor theories, the most general disformal transformation leaves $\varphi$ invariant, but replaces $g_{ab}$ by $g'_{ab}$ where

$$g'_{ab} := A(\varphi,\rho) g_{ab} + B(\varphi,\rho) \varphi_a \varphi_b , \qquad \text{Eq.3.1}$$

where A and B are differentiable functions of $\varphi$ and $\rho$. To assure that $g'_{ab}$ is of the same signature as $g_{ab}$ it is required that

$$A + \rho B > 0 . \qquad \text{Eq.3.2}$$

When $B = 0$, the transformation presented in Eq.3.1 is a special case of a conformal transformation, which we know leaves P invariant. We shall just examine Eq.3.1 for the case where $A = 1$, and to simplify matters even further, we shall let $B = 1$. Thus we are going to consider the transformation

$$g_{ab} \rightarrow g'_{ab} := g_{ab} + \varphi_a \varphi_b , \qquad \text{Eq.3.3}$$

where Eq.3.2 requires that

$$1 + \rho > 0 . \qquad \text{Eq.3.4}$$

Henceforth, whenever I refer to a disformal transformation in this paper, I will be referring to a transformation of the form given by Eq.3.3, with ρ satisfying Eq.3.4.

A lengthy, but straightforward calculation, shows that under a disformal transformation

$$P' = P + [4\varepsilon^{abjl} \varphi^h \varphi_{ij} R_h{}^i{}_{ab} / (1+\rho)]_{|l}, \qquad \text{Eq.3.5}$$

where P' means P built from the $g'_{ab}$ metric. Thus we see that P' equals P plus a divergence.

Recall that our original motivation in this paper was using Lagrange multipliers to perform constrained variations of metric Lagrangians. When the constraint is P = 0, we add φP to our original Lagrangian. So let us now look at the effect of a disformal transformation on φP.

Upon multiplying Eq.3.5 by φ we get

$$\varphi P' = \varphi P + [4\varphi\, \varepsilon^{abjl}\varphi^h \varphi_{ij} R_h{}^i{}_{ab} / (1+\rho)]_{|l} - 4\varepsilon^{abjl}\varphi^h \varphi_l \varphi_{ij} R_{h\,iab} / (1+\rho). \qquad \text{Eq.3.6}$$

Consequently we see that φP is not invariant under a disformal transformation of the form given in Eq.3.3, but gives rise to a new class of Lagrangians

$$L_{PD} := K\, \varepsilon^{rstu} \varphi_u\, \varphi^p\, \varphi_{qt}\, R_p{}^q{}_{rs} \qquad \text{Eq.3.7}$$

where K = K(φ,ρ) is a differentiable real valued function. I call this Lagrangian $L_{PD}$ since it can be obtained from $L_P$ by means of a disformal transformation. Since $L_{PD}$ differs from the Lagrangians I presented in [2], we know that it must yield EL tensors which are greater than second order, and perhaps of fourth order. To see that these EL tensors are actually of third order we do not really need to compute $E^{ab}(L_{PD})$ and $E(L_{PD})$ in their entirety, but only their possible fourth order terms. This can be accomplished using some of the machinery presented in [2], which permits us to express the EL tensors as a sum of tensorial quantities. Eqs.4.8-4.11 in [2] tell us that for a Lagrangian of the form $L = L(g_{ij}\,;\,g_{ij,h}\,;\,g_{ij,hk}\,;\,\varphi\,;\,\varphi_{,h};\,\varphi_{,hk})$

$$E^{ab}(L) = -\Pi^{ab,hk}{}_{|hk} + \Pi^{ab,h}{}_{|h} - \Pi^{ab} \qquad \text{Eq.3.8}$$

and

$$E(L) = -\zeta^{hk}{}_{|hk} + \zeta^{h}{}_{|h} - \zeta \qquad \text{Eq.3.9}$$

where

$$\Pi^{ab,hk} := \frac{\partial L}{\partial g_{ab,hk}}\,;\quad \zeta^{hk} := \frac{\partial L}{\partial \varphi_{,hk}}\,;\quad \zeta := \frac{\partial L}{\partial \varphi}\,; \qquad \text{Eq.3.10}$$

$$\Pi^{ab,h} = \tfrac{1}{2}(\zeta^{ab}\,\varphi^{h} - \zeta^{hb}\,\varphi^{a} - \zeta^{ha}\,\varphi^{b})\,; \qquad \text{Eq.3.11}$$

$$\Pi^{ab} = \tfrac{1}{3}R_{k}{}^{b}{}_{mh}\Pi^{hk,am} - R_{k}{}^{a}{}_{mh}\Pi^{hk,bm} - \tfrac{1}{2}\varphi^{a}\,\zeta^{b} - \zeta^{bl}\,\varphi_{l}{}^{a} + \tfrac{1}{2}g^{ab}\,L \qquad \text{Eq.3.12}$$

and

$$\zeta^{a} := \frac{\partial L}{\partial \varphi_{,a}} + \zeta^{rs}\,\Gamma^{a}{}_{rs}, \qquad \text{Eq.3.13}$$

with $\Gamma^{a}{}_{rs}$ denoting the components of the Christoffel symbols of the second kind. (Note that in Eq.4.11 of [2] there is a typographical error in the equation for $\Pi^{ab}$ involving $\varphi^{a}\,\zeta^{b}$, which is corrected in Eq.3.12 above.) Using Eq.3.7 it is easily seen

that

$$\Pi^{ab,hk} = \frac{\partial L_{PD}}{\partial g_{ab,hk}} = \tfrac{1}{4} K\{^{abhk}\} \qquad \text{Eq.3.14}$$

where

$$\{^{abhk}\} := \varepsilon^{kbtu}\varphi_u \varphi^a \varphi^h_t + \varepsilon^{katu}\varphi_u \varphi^b \varphi^h_t + \varepsilon^{hbtu}\varphi_u\varphi^a \varphi^k_t + \varepsilon^{hatu}\varphi_u \varphi^b \varphi^k_t +$$
$$+ \varepsilon^{bktu}\varphi_u \varphi^h\varphi^a_t + \varepsilon^{aktu}\varphi_u \varphi^h \varphi^b_t + \varepsilon^{bhtu}\varphi_u \varphi^k \varphi^a_t + \varepsilon^{ahtu}\varphi_u \varphi^k \varphi^b_t \qquad \text{Eq.3.15}$$

and

$$\zeta^{ab} = \tfrac{1}{2}K(\varepsilon^{rsbu}\varphi_u\varphi^p R_p{}^a{}_{rs} + \varepsilon^{rsau}\varphi_u\varphi^p R_p{}^b{}_{rs}). \qquad \text{Eq.3.16}$$

Eqs.3.11 and 3.16 allow us to conclude that

$$\Pi^{ab,h} = \tfrac{1}{4}K(\varepsilon^{rsbu}\varphi^h\varphi_u\varphi^p R_p{}^a{}_{rs} + \varepsilon^{rsau}\varphi^h\varphi_u\varphi^p R_p{}^b{}_{rs} - \varepsilon^{rsbu}\varphi^a\varphi_u\varphi^p R_p{}^h{}_{rs} +$$
$$- \varepsilon^{rshu}\varphi^a\varphi_u\varphi^p R_p{}^b{}_{rs} - \varepsilon^{rsau}\varphi^b\varphi_u\varphi^p R_p{}^h{}_{rs} - \varepsilon^{rshu}\varphi^b\varphi_u\varphi^p R_p{}^a{}_{rs}). \qquad \text{Eq.3.17}$$

There are three more quantities which need to be computed in order to determine $E^{ab}(L_{PD})$ and $E(L_{PD})$; viz., $\zeta$, $\zeta^a$ and $\Pi^{ab}$. $\zeta$ is obviously given by

$$\zeta = K_\varphi \, \varepsilon^{rstu} \varphi_u \varphi^p \varphi_{qt} R_p{}^q{}_{rs}, \qquad \text{Eq.3.18}$$

while, with a bit more work, we get

$$\zeta^a = 2K_\rho \varphi^a \varepsilon^{rstu} \varphi_u \varphi^p \varphi_{qt} R_p{}^q{}_{rs} + K\varepsilon^{rsta} \varphi^p \varphi_{qt} R_p{}^q{}_{rs} + K\varepsilon^{rstu} \varphi_u \varphi_{qt} R^{aq}{}_{rs}. \qquad \text{Eq.3.19}$$

Using Eqs.3.12, 3.14, 3.15, 3.16 and 3.19 we find that

$$\Pi^{ab} = \tfrac{1}{12} K R_k{}^b{}_{mh} \{^{hkam}\} - \tfrac{1}{4}K R_k{}^a{}_{mh} \{^{hkbm}\} - K_\rho \varphi^a \varphi^b \varepsilon^{rstu} \varphi_u \varphi^p \varphi_{qt} R_p{}^q{}_{rs} +$$
$$- \tfrac{1}{2}K \varepsilon^{rstb} \varphi^a \varphi^p \varphi_{qt} R_p{}^q{}_{rs} - \tfrac{1}{2}K \varepsilon^{rstu} \varphi^a \varphi_u \varphi_{qt} R^{bq}{}_{rs} - \tfrac{1}{2}K \varepsilon^{rsbu}\varphi_u \varphi^p \varphi^a_l R_p{}^l{}_{rs} +$$
$$- \tfrac{1}{2}K \varepsilon^{rslu} \varphi_u \varphi^p \varphi^a_l R_p{}^b{}_{rs} + \tfrac{1}{2}K g^{ab}\varepsilon^{rstu} \varphi_u \varphi^p \varphi_{qt} R_p{}^q{}_{rs}. \qquad \text{Eq.3.20}$$

We could now employ Eqs.3.8, 3.9, 3.14 and 3.16-3.20 to obtain expressions for $E^{ab}(L_{PD})$ and $E(L_{PD})$. I am not going to bother to write these EL equations out. In a moment we shall look at the fourth order terms in $E^{ab}(L_{PD})$ and $E(L_{PD})$, to make sure that they vanish. But before doing that let us examine the trace of $E^{ab}(L_{PD})$. From Eq.3.8 we see that

$$g_{ab} E^{ab}(L_{PD}) = -(g_{ab} \Pi^{ab,hk})_{|hk} + (g_{ab} \Pi^{ab,h})_{|h} - g_{ab} \Pi^{ab} . \qquad \text{Eq.3.21}$$

Thanks to Eqs.3.14, 3.15. 3.17 and the first and second Bianchi identities

$$g_{ab} E^{ab}(L_{PD}) = -g_{ab} \Pi^{ab} = -(K + \rho K_\rho)\varepsilon^{rstu} \varphi_u \varphi^p \varphi_{qt} R_p{}^q{}_{rs} .$$

Thus $g_{ab} E(L_{PD}) = 0$, if and only if $K = k(\varphi)/\rho$, where $k(\varphi)$ is an arbitrary differentiable function of $\varphi$. As a result the Lagrangian

$$L_{PC2} := k(\varphi) \varepsilon^{rstu} \varphi_u \varphi^p \varphi_{qt} R_p{}^q{}_{rs}/\rho \qquad \text{Eq.3.22}$$

is such that

$$g_{ab} E^{ab}(L_{PC2}) = 0 . \qquad \text{Eq.3.23}$$

We can now use **Prop.2.1** to conclude that $L_{PC2}$ must be conformally invariant up to a divergence, and hence $E^a{}_b(L_{PC2})$ is conformally invariant. If you do not wish to use **Prop.2.1**, then a direct calculation will show that $L_{PC2}$ is conformally invariant, with no divergence hanging around.

Let us now turn our attention to the fourth order terms in $E^{ab}(L_{PD})$. From Eqs. 3.8, 3.14 and 3.17 it is apparent that there are no fourth order $g_{ab}$ terms in $E^{ab}(L_{PD})$, and that the fourth order φ terms are all in $\Pi^{ab,hk}{}_{|hk}$. These fourth order terms are

$$4^{th} \text{ order } E^{ab}(L_{PD}) = \tfrac{1}{4}K\{\varepsilon^{kbtu}\varphi_u\varphi^a\varphi^h{}_{thk} + \varepsilon^{katu}\varphi_u\varphi^b\varphi^h{}_{thk} + \varepsilon^{hbtu}\varphi_u\varphi^a\varphi^k{}_{thk} +$$

$$+ \varepsilon^{hatu}\varphi_u\varphi^b\varphi^k{}_{thk} + \varepsilon^{bktu}\varphi_u\varphi^h\varphi^a{}_{thk} + \varepsilon^{aktu}\varphi_u\varphi^h\varphi^b{}_{thk} + \varepsilon^{bhtu}\varphi_u\varphi^k\varphi^a{}_{thk} +$$

$$+ \varepsilon^{ahtu}\varphi_u\varphi^k\varphi^b{}_{thk}\} . \qquad \text{Eq.3.24}$$

I shall now show you how the fourth order terms in φ disappear in Eq.3.24. The first and third terms in Eq.3.24 are given by ( note that I have interchanged h and k in the third term)

$$\tfrac{1}{4}K\{\varepsilon^{kbtu}\varphi_u\varphi^a\varphi^h{}_{thk} + \varepsilon^{kbtu}\varphi_u\varphi^a\varphi^h{}_{tkh}\} = \tfrac{1}{4}K\{\varepsilon^{kbtu}(\varphi_u\varphi^a\varphi^h{}_{tkh} + \varphi_u\varphi^a\varphi^m{}_t R_m{}^h{}_{hk} +$$

$$- \varphi_u\varphi^a\varphi^h{}_m R_t{}^m{}_{hk}) + \varepsilon^{kbtu}\varphi_u\varphi^a\varphi^h{}_{tkh}\} . \qquad \text{Eq.3.25}$$

However,

$$\varepsilon^{kbtu}\varphi^h{}_{tkh} = \tfrac{1}{2}\varepsilon^{kbtu}(\varphi^l R_l{}^h{}_{tk})_{|h}$$

which is, at most, third order in $g_{ab}$ and second order in φ. As a result we can use Eq.3.25 to deduce that the first and third terms on the right-hand side of Eq.3.24 reduce to third order in $g_{ab}$ and second order in φ. A similar analysis can be applied to three other pairs of terms in Eq.3.24 to show that the fourth order terms in $E^{ab}(L_{PD})$ vanish. In fact, due to Eqs.3.8 and 3.17 we can now conclude that $E^{ab}(L_{PD})$ is at most third order in both $g_{ab}$ and φ.

We shall now turn our attention to $E(L_{PD})$. Using Eqs. 3.9, 3.16, 3.18 and 3.19 we see that the only fourth order terms in $E(L_{PD})$ are 4th order in $E(L_{PD}) = -½K (\varepsilon^{rsbu}\varphi_u \varphi^p R_p{}^a{}_{rs|ab} + \varepsilon^{rsau} \varphi_u \varphi^p R_p{}^b{}_{rs|ab})$ .    Eq.3.26

Evidently

$$\varepsilon^{rsau} R_p{}^b{}_{rs|ab} = 0 ,$$

due to the second Bianchi identity, while

$$\varepsilon^{rsbu} R_p{}^a{}_{rs|ab} = \varepsilon^{rsbu} R_p{}^a{}_{rs|ba} + 2^{nd} \text{ order terms} = 2^{nd} \text{ order terms.}$$

Consequently, $E(L_{PD})$ is devoid of fourth order terms and is in fact at most third order in both $g_{ab}$ and $\varphi$.

We began this section examining the effects of the disformal transformation on the Pontrjagin Lagrangian, P. This led to the discovery of the Lagrangian $L_{PD}$, given in Eq.3.7. If we perform a disformal transformation of this Lagrangian then do we arrive at another new class of Lagrangians? The answer is no. To see that this is so, let us consider the Lagrangian

$$L_{PD'} := k\, \varepsilon^{rstu} \varphi_u \varphi^p \varphi_{qt} R_p{}^q{}_{rs}/ \rho^2 ,$$    Eq.3.27

where k is an arbitrary function of $\varphi$. It is easily seen that $L_{PD'}$ is invariant under a disformal transformation. As a result the class of Lagrangians $L_{PD}$ would be mapped into itself by disformal transformations. Consequently we now have two classes of Lagrangians, $L_P$ and $L_{PD}$ which are related by disformal transformations, and produce EL equations which are at most of third order.

I shall now make a few remarks about the effects of disformal transformations on the class of Gauss-Bonnet Lagrangians $L_{GB}$.

It is a straight-forward, albeit lengthy, matter to show that under a disformal transformation

$$GB' = GB/(1+\rho)^{1/2} - 4(g)^{1/2}\delta^{hijk}{}_{rstu}\varphi^r{}_h\,\varphi^s{}_i\,R^{tu}{}_{jk}/(1+\rho)^{3/2} +$$

$$+ 4(g)^{1/2}\delta^{hijk}{}_{rstu}\varphi^r{}_h\,\varphi^s{}_i\,\varphi^t{}_j\,\varphi^u{}_k/(1+\rho)^{5/2}\ .\qquad\text{Eq.3.28}$$

In deriving this result I needed to make use of several dimensionally dependent identities akin to those presented by Lovelock in [31]. The identities that I required were

$$0 = \delta^{abcde}{}_{hijkl}\varphi^h\,\varphi_a\,R^{ij}{}_{bc}\,R^{kl}{}_{de}\,,\ \ 0 = \delta^{abcde}{}_{hijkl}\varphi^h\,\varphi_a\,\varphi^i{}_b\,\varphi^j{}_c\,R^{kl}{}_{de}$$

and

$$0 = \delta^{abcde}{}_{hijkl}\varphi^h\,\varphi_a\,\varphi^i{}_b\,\varphi^j{}_c\,\varphi^k{}_d\,\varphi^l{}_e\,.$$

Eq.3.28 leads us to consider two more second order scalar-tensor Lagrangians, *viz.*,

$$L_{GBD1} := V\,(g)^{1/2}\,\delta^{hijk}{}_{rstu}\,\varphi^r{}_h\,\varphi^s{}_i\,R^{tu}{}_{jk}\qquad\text{Eq.3.29}$$

and

$$L_{GBD2} := W\,(g)^{1/2}\,\delta^{hijk}{}_{rstu}\,\varphi^r{}_h\,\varphi^s{}_i\,\varphi^t{}_j\,\varphi^u{}_k\qquad\text{Eq.3.30}$$

where V and W are differentiable functions of $\varphi$ and $\rho$. From [2] we see that these Lagrangians are not among those that yield second order EL equations, and so they yield EL equations of either third or fourth order. Using an approach analogous to the one we used to analyze the EL equations of $L_{PD}$, it is possible to show that both $L_{GBD1}$ and $L_{GBD2}$ yield third order EL equations. So now we have five Lagrangians that yield third order scalar-tensor field equations. However, can we generate even more by applying a disformal transformation to $L_{GBD1}$ and $L_{GBD2}$? Fortunately not, since a disformal transformation of a Lagrangian of the form $L_{GBD1}$, produces a Lagrangian which is the sum of $L_{GBD1}$ and $L_{GBD2}$ type Lagrangians. While a disformal transformation leaves the class of $L_{GBD2}$ type Lagrangians invariant. In fact the Lagrangian

$$L_{GBD2'} := w\, (g)^{½}\, \delta^{hijk}{}_{rstu}\, \varphi^{rh}\, \varphi^s{}_i\, \varphi^t{}_j\, \varphi^u{}_k / (\rho)^{9/2} \qquad \text{Eq.3.31}$$

is disformally invariant, where w is an arbitrary differentiable function of φ. Moreover, none of the $L_{GB}$, $L_{GBD1}$ and $L_{GBD2}$ classes of Lagrangians admit conformally invariant subclasses.

**Section 4: Scalar-Tensor-Connection Field Theories**

The two Lagrangians which essentially are the basis of this paper are GB and P. However, P, unlike GB, can be built without using a metric tensor. To see this we shall rewrite it as

$$P = -\,\varepsilon^{abcd}\, R_l{}^m{}_{ab}\, R_m{}^l{}_{cd}\,.$$

Let $[{}_{rs}{}^t]$ denote the local components of an arbitrary affine connection [ ] (read: square bracket), on our orientable 4-manifold. The Pontrjagin Lagrangian of [ ] would then be given by

$$P_{[\,]} := -\varepsilon^{abcd} K^m{}_{l\,ab} K^l{}_{m\,cd} \qquad \text{Eq.4.1}$$

where the curvature tensor components of [ ] are

$$K^m{}_{l\,ab} := [{}_{al}{}^m]_{,b} - [{}_{bl}{}^m]_{,a} + [{}_{al}{}^p][{}_{bp}{}^m] - [{}_{bl}{}^p][{}_{ap}{}^m].$$

So we see that $P_{[\,]}$ is independent of the metric tensor. We now consider the Lagrangian

$$L_{P[\,]} := M\, P_{[\,]} \qquad \text{Eq.4.2}$$

where M is an arbitrary differentiable function of $\varphi$ and $\rho$. So $L_{P[\,]}$ is a Lagrangian which is of first order in $\varphi$, zeroth order in $g_{ab}$ and first order in $[{}_{rs}{}^t]$. If $L = L(g_{ab}; \varphi; \varphi_{,a}; [{}_{rs}{}^t]; [{}_{rs}{}^t]_{,u})$, then its associated EL tensors are defined by

$$E^{rs}{}_t(L) := \frac{d}{dx^u}\frac{\partial L}{\partial [{}_{rs}{}^t]_{,u}} - \frac{\partial L}{\partial [{}_{rs}{}^t]} \qquad \text{Eq.4.3}$$

$$E^{rs}(L) := -\frac{\partial L}{\partial g_{rs}} \qquad \text{Eq.4.4}$$

and

$$E(L) := \frac{d}{dx^u}\frac{\partial L}{\partial \varphi_{,u}} - \frac{\partial L}{\partial \varphi} \qquad \text{Eq.4.5}$$

A straightforward calculation using $L_{P[\,]}$ in Eqs.4.3-4.5 shows that

$$E^{rs}{}_t(L_{P[\,]}) = -4\, M_{,u}\varepsilon^{rucd} K^s{}_{t\,cd} - 4\, M\, \varepsilon^{rucd} K^s{}_{t\,cd;u} + 4\, M\, \varepsilon^{rucd} K^s{}_{t\,ld} S_{uc}{}^l \qquad \text{Eq.4.6}$$

$$E^{rs}(L_{P[\,]}) = M_\rho\, \varphi^r\, \varphi^s\, \varepsilon^{abcd}\, K^m{}_{l\,ab}\, K^l{}_{m\,cd}, \qquad \text{Eq.4.7}$$

and

$$E(L_{P[\,]}) = 2\frac{d}{dx^u}(M_\rho \varphi^u P) - M_\varphi P \qquad \text{Eq.4.8}$$

where $S_{uc}{}^l := [{}_{uc}{}^l] - [{}_{cu}{}^l]$, is the torsion tensor of [ ]. If $T^{ij}$ is a tensor density of weight w, then the components of its covariant derivative with respect to [ ] are given by

$$T^i{}_{j;k} := T^{ij}{}_{,k} + T^p{}_j [{}_{kp}{}^i] - T^i{}_p [{}_{kj}{}^p] - w\, T^i{}_j [{}_{pk}{}^p]. \qquad \text{Eq.4.9}$$

Eq.4.6 can be simplified due to the second Bianchi identity for $K_l{}^m{}_{ab}$ (*see*, [20], noting that their definition of covariant differentiation differs from the one given in Eq.4.9), which implies that

$$\varepsilon^{rucd}(K_t{}^s{}_{cd;u} - K_t{}^s{}_{ld}\, S_{uc}{}^l) = 0.$$

As a result Eq.4.6 reduces to

$$E^{rs}{}_t(L_{P[\,]}) = -4\, M_{,u}\, \varepsilon^{rucd}\, K_{ts\,cd}. \qquad \text{Eq.4.10}$$

First we should note that if M is a constant then $E^{rs}{}_t(L_{P[\,]}) = 0$, and so the EL tensor of $P_{[\,]}$ vanishes, just as it does in the metric case. (This fact is used in [25] to show that $P_{[\,]}$ can be expressed as a divergence using an affine connection and local parallelization.) If $M_\rho \neq 0$, then the EL equations corresponding to Eqs.4.7 and 4.10 would imply that

$$P_{[\,]} = 0, \quad \text{and} \quad M_{,u}\, \varepsilon^{rucd}\, K_t{}^s{}_{cd} = 0. \qquad \text{Eq.4.11}$$

Thus these EL equations would be at most first order in $g_{ab}$ and $[{}_{rs}{}^t]$, and second order in $\varphi$. If $M_\rho = 0$, and $M_\varphi \neq 0$, then we still arrive at Eq.4.11 as the field equation, with $M_{,u} = M_\varphi \, \varphi_{,u}$, and $P_{[\,]}$ must vanish, due to the EL equation we get from Eq.4.8. Thus when $M_\rho = 0$, the field equation are first order in $\varphi$ and $[_{rs}{}^t]$, and they place no restriction on $g_{ab}$ since it is absent from the equations. In either case, as long as M is not a constant, the field equation imply that $P_{[\,]} = 0$.

The above work shows that, as far as differential order is concerned, this system of scalar-metric-connection equations is simpler then their scalar-tensor counterparts based on the Lagrangian $L_P$ given in Eq.2.2. However, this simplification has been obtained at a price. $[_{rs}{}^t]$ has 64 independent components in a 4-space, and hence Eq.4.11 involves 75 variables. If $[\,]$ is a symmetric connection, then $[_{rs}{}^t]$, would only have 40 independent components, and the system of equations given in Eq.4.11 would have 51 variables. In either case, this is far more than the 11 variables in the EL Equations generated by $L_P$ in the scalar-tensor theory. It is amusing to note that if $[_{rs}{}^t]$ is symmetric then it is possible to do a constrained variation of the Lagrangian $L_{P[\,]}$ to obtain the exact same equations we generated with $L_P$ in the scalar-tensor theory. For that we would consider the Lagrangian

$$L_{P[\,]\text{constrained}} := L_{P[\,]} + \Lambda^{rst}(g_{rs,t} - g_{ls}[_{tr}{}^l] - g_{rl}[_{ts}{}^l]),$$

where the tensor density $\Lambda^{rst} \equiv \Lambda^{srt}$ is our Lagrange undetermined multiplier. The EL equation $E_{rst}(L_{P[\,]\text{constrained}}) := -\partial L_{P[\,]\text{constrained}}/\partial \Lambda^{rst} = 0$, implies that the symmetric connection must be the Levi-Civita connection. With a little effort it is easy to see that the remaining EL equations give rise to the scalar-tensor equations generated by $L_P$.

Dealing with the scalar-tensor-connection counterpart of the Lagrangian $L_{PD}$ would be more difficult than our handling of $L_{P[\,]}$. This is so because two possibilities would arise in this case. In the first we could use the Levi-Civita connection to deal with the covariant derivative of $\varphi$, and in the second we could take covariant derivatives of $\varphi$ with respect to [ ]. In the latter case it must be noted that $\varphi_{;ab} \neq \varphi_{;ba}$ if [ ] is not symmetric. I leave it to the interested reader to pursue these Lagrangians further.

What we did with the Lagrangian $L_{P[\,]}$ is akin to the Palatini formalism described in the context of f(R) theories by Sotiriou and Faraoni [19]. But here we are applying these ideas to scalar-tensor Lagrangians.

At this point it might be interesting to examine the Gauss-Bonnet Lagrangian, when it is viewed as a concomitant of $g_{ab}$ and $[{}_{rs}^{t}]$. The scalar-metric-connection Lagrangian for that problem would be

$$L_{GB[\,]} := N\, GB_{[\,]},$$

where N is a differentiable function of $\varphi$ and $\rho$, and

$$GB_{[\,]} := \tfrac{1}{4}\, (g)^{\frac{1}{2}}\, \delta^{abcd}{}_{hijk}\, g^{lh}\, K^{i}{}_{l\,ab}\, g^{mj}\, K_{m}{}^{k}{}_{cd}.$$

The EL tensors of $L_{GB[\,]}$ can be computed using Eqs.4.3-4.5. I leave it to the reader to determine their explicit form. It suffices to say that in general the system will be at most 2$^{nd}$ order in $\varphi$ and $[_{rs}{}^t]$, and first order in $g_{ab}$.

If one wishes to explore Lagrangians of the form

$$L = L(g_{ab}; g_{ab,c};...; \varphi; \varphi_{,a}; ...; [_{rs}{}^t]; [_{rs}{}^t]_{,u}; ...),$$

then there is an interesting identity relating the associated EL tensors. The identity can be found in [22] (*see,* Eq.5.34 in that paper) and is given by

$$2\frac{\partial}{\partial x^t}(E^{tb}(L)g_{rb}) - E^{ab}(L)g_{ab,r} - E(L)\varphi_{,r} + \frac{\partial}{\partial x^t}(E^{tb}{}_k(L)[_{rb}{}^k]) + \frac{\partial}{\partial x^t}(E^{bt}{}_k(L)[_{br}{}^k]) +$$

$$- \frac{\partial}{\partial x^t}(E^{ab}{}_r(L)[_{ab}{}^t]) - \frac{\partial^2}{\partial x^t \partial x^j}E^{tj}{}_r(L) - E^{ab}{}_c(L)[_{ab}{}^c]_{,r} = 0 .$$

This identity can be rewritten in a manifestly tensorial form as follows:

$$2E_r{}^b(L)_{|b} - \varphi_{,r} E(L) - E^{ab}{}_r(L)_{;ab} + E^{ab}{}_c(L)_{;a} S_{rb}{}^c - E^{ab}{}_c(L) K_b{}^c{}_{ar} = 0 . \qquad \text{Eq.4.12}$$

Eq.4.12 implies that $E(L) = 0$, if $E^{ab}(L)$ and $E^{ab}{}_c(L) = 0$. It also implies that $E^{ab}(L)$ is divergence-free, when $E(L) = 0$ and $E^{ab}{}_c(L) = 0$. This is useful in the presence of matter, since in that case $E^{ab}(L)$ is usually set equal to $T_M{}^{ab}$, the energy-momentum tensor of matter, which will then be divergence-free (with respect to the Levi-Civita connection) when the EL equations hold.

**Section 5: Summary and Concluding Remarks**

I began this paper with an examination of the variation of the Lagrangian

$(g)^{½} R_{hijk} R^{hijk}$, subject to the constraint that $GB = 0$, for all metrics through which we vary $(g)^{½} R_{hijk} R^{hijk}$. This in turn led us to consider Lagrangians involving the Pontjagin Lagrangian, P. The first such Lagrangian was

$$L_P := J\, P, \qquad\qquad\qquad\qquad\qquad\qquad\qquad\text{Eq.5.1}$$

where J is a differentiable function of φ and ρ. This Lagrangian yielded the EL tensors presented in Eqs.2.4 and 2.7, which were at most of third order in the derivatives of φ and $g_{ab}$. We also found a subclass of the $L_P$ Lagrangians, $L_{PC}$, which is conformally invariant, where

$$L_{PC} := j(\varphi)\, P. \qquad\qquad\qquad\qquad\qquad\qquad\text{Eq.5.2}$$

If we were not going to couple $L_{PC}$ to another Lagrangian involving φ, then there would be no loss of generality if we chose $j(\varphi) = \alpha\, \varphi$ in Eq.5.2, where α is just a coupling constant. By examining the effect of the disformal transformation

$$g_{ab} \to g'_{ab} := g_{ab} + \varphi_a\, \varphi_b$$

on the Lagrangian $L_P$ we were led to a second class of Lagrangians

$$L_{PD} := K\, \varepsilon^{rstu}\, \varphi_u\, \varphi^p\, \varphi_{qt}\, R_p{}^q{}_{rs}\,, \qquad\qquad\qquad\text{Eq.5.3}$$

where K is a differentiable function of φ and ρ. The EL equations of $L_{PD}$ are far more complex than those for $L_P$, but nevertheless, are at most of third order in the derivatives of φ and $g_{ab}$. We saw that the class of Lagrangians, $L_{PD}$, is mapped into itself under disformal transformations. The $L_{PD}$ class contains subclasses, $L_{PC2}$ and

$L_{PD'}$, defined by

$$L_{PC2} := k\, \varepsilon^{rstu}\, \varphi_u\, \varphi^p\, \varphi_{qt}\, R_p{}^q{}_{rs} / \rho \qquad \text{Eq.5.4}$$

and

$$L_{PD'} := k\, \varepsilon^{rstu}\, \varphi_u\, \varphi^p\, \varphi_{qt}\, R_p{}^q{}_{rs} / \rho^2 \qquad \text{Eq.5.5}$$

where k is an arbitrary differentiable function of $\varphi$. $L_{PC2}$ is invariant under a general conformal transformation, while $L_{PD'}$ is invariant under a disformal transformation.

In **Section 3** we also presented three other classes of Lagrangians, two of which were generated from the Gauss-Bonnet Lagrangian, using disformal transformations. These three Lagrangians were

$$L_{GB} := U\, GB, \qquad \text{Eq.5.6}$$

$$L_{GBD1} := V\, (g)^{½}\, \delta^{hijk}{}_{rstu}\, \varphi^r_h\, \varphi^s_i\, R^{tu}{}_{jk} \qquad \text{Eq.5.7}$$

and

$$L_{GBD2} := W\, (g)^{½}\, \delta^{hijk}{}_{rstu}\, \varphi^r_h\, \varphi^s_i\, \varphi^t_j\, \varphi^u_k \qquad \text{Eq.5.8}$$

where U, V and W are differentiable functions of $\varphi$ and $\rho$. Like $L_P$ and $L_{PD}$, each of these three classes of second order Lagrangians, produce EL equations which are at most of third order. None of the Lagrangians presented in Eqs.5.6-5.8 admit conformally invariant subclasses, but $L_{GBD2}$ admits a subclass of disformally invariant Lagrangians. This subclass is given by

$$L_{GBD2'} := w\, (g)^{½}\, \delta^{hijk}{}_{rstu}\, \varphi^r_h\, \varphi^s_i\, \varphi^t_j\, \varphi^u_k / (\rho)^{9/2}$$

where w is an arbitrary differentiable function of φ.

The five classes of second order Lagrangians presented above ostensibly differ from the two other classes of second order Lagrangians presented in [5] and [7] which also yield third order scalar-tensor EL equations. These two other Lagrangians are given by

$$^{BH}L_4 := F_4 (g)^{1/2} \delta^{hij}{}_{rst} \varphi^r \varphi_h \varphi^s{}_i \varphi^t{}_j \qquad \text{Eq.5.9}$$

and

$$^{BH}L_5 := F_5 (g)^{1/2} \delta^{hijk}{}_{rstu} \varphi^r \varphi_h \varphi^s{}_i \varphi^{tj} \varphi^u{}_k \qquad \text{Eq.5.10}$$

where $F_4$ and $F_5$ are arbitrary differentiable functions of φ and ρ.

Lagrangian $^{BH}L_5$ certainly looks similar to $L_{GBD2}$ given in Eq.5.8. So let us take a closer look. Using the formula for anti-symmetrized second covariant derivatives, along with the second Bianchi identity, we find (where I have dropped the subscript 5 of off $F_5$)

$$(g)^{1/2} F \delta^{hijk}{}_{rstu} \varphi^r \varphi_h \varphi^s{}_i \varphi^t{}_j \varphi^u{}_k = [(g)^{1/2} F \delta^{hijk}{}_{rstu} \varphi^r \varphi \varphi^s{}_i \varphi^t{}_j \varphi^u{}_k]_{|h} +$$

$$-(g)^{1/2} \varphi F_\varphi \delta^{hijk}{}_{rstu} \varphi^r \varphi_h \varphi^s{}_i \varphi^t{}_j \varphi^u{}_k - 2(g)^{1/2} \varphi F_\rho \delta^{hijk}{}_{rstu} \varphi^r \varphi_l \varphi^l{}_h \varphi^s{}_i \varphi^t{}_j \varphi^u{}_k +$$

$$-(g)^{1/2} \varphi F \delta^{hijk}{}_{rstu} \varphi^r{}_h \varphi^s{}_i \varphi^t{}_j \varphi^u{}_k + {}^3/_2 (g)^{1/2} \varphi F \delta^{hijk}{}_{rstu} \varphi^r \varphi_l R^{ls}{}_{hi} \varphi^t{}_j \varphi^u{}_k. \qquad \text{Eq.5.11}$$

The second term on the right-hand side of Eq.5.11 can be taken over to the left side of the equation to give us two terms of the form $^{BH}L_5$, while the fourth term on the right-hand side of Eq.5.11 is of the form of the Lagrangian $L_{GBD2}$. Using the identity

$$0 = \delta^{abcde}{}_{hijkl}\, \varphi^h\, \varphi_a\, \varphi^i{}_b\, \varphi^j{}_c\, \varphi^k{}_d\, \varphi^l{}_e$$

we see (upon expanding the generalized Kronecker delta about row a) that the third term on the right-hand side of Eq.5.11 is equal to

$$-\tfrac{1}{2}(g)^{1/2}\varphi\rho F_\rho\, \delta^{hijk}{}_{rstu}\, \varphi^{rh}\, \varphi^s{}_i\, \varphi^t{}_j\, \varphi^u{}_k ,$$

and thus the third term on the right-hand side of Eq.5.11 is also of the form of the Lagrangian $L_{GBD2}$. What about the last term on the right-hand side of Eq.5.11? Using the identity

$$0 = \delta^{abcde}{}_{hijkl}\, \varphi^h\, \varphi_a\, \varphi^i{}_b\, \varphi^j{}_c\, R^{kl}{}_{de}$$

we find (once again expanding the generalized Kronecker delta about the index a) that this term is equal to

$$\tfrac{3}{4}(g)^{1/2}\varphi\rho F\, \delta^{hijk}{}_{rstu}\, \varphi^r{}_h\, \varphi^s{}_i\, R^{tu}{}_{jk} - \tfrac{3}{4}(g)^{1/2}\varphi F\, \delta^{hijk}{}_{rstu}\, \varphi^r\, \rho_{|h}\, \varphi^s{}_i\, R^{tu}{}_{jk} .$$

So the last term on the right-hand side of Eq.5.11 is of the form of the Lagrangian $L_{GBD1}$ added to the Lagrangian

$$-\tfrac{3}{4}(g)^{1/2}\varphi F\, \delta^{hijk}{}_{rstu}\, \varphi^r\, \rho_{|h}\, \varphi^s{}_i\, R^{tu}{}_{jk} .$$

Now we can "pull the index h" out of the above Lagrangian to rewrite it as a sum of an $L_{GB}$, and $L_{GBD1}$ Lagrangian, with a Lagrangian of the form

$$L = (g)^{1/2} F\, \delta^{hijk}{}_{rstu}\, \varphi^r\, \varphi_h\, \varphi^s{}_i\, R^{tu}{}_{jk} \qquad \text{Eq.5.12}$$

where $F$ is a differentiable function of $\varphi$ and $\rho$. I have been unable to further simplify $L$ into a sum of Lagrangians of the form $L_{GB}$, $L_{GBD1}$ and $L_{GBD2}$. So, due to the above work, we see that the Lagrangian $L$ and $^{BH}L_5$ differ by a divergence, and Lagrangians of the form $L_{GB}$, $L_{GBD1}$ and $L_{GBD2}$. Thus $L$ does not really represent anything new.

If you look at the four Lagrangians used in Horndeski Scalar Theory, you will note that two of them, conventionally denoted by $L_4$ and $L_5$, are composite Lagrangians. We shall let $L_{4L}$ and $L_{4R}$ denote the left and right parts of $L_4$, and $L_{5L}$ and $L_{5R}$ denote the left and right parts of $L_5$. If we use either half of the pair of Lagrangians which make up $L_4$ and $L_5$, we would obtain second order Lagrangians, which yield third order EL equations, with the third order parts cancelling when the Lagrangian pairs are combined. This leads to two more Lagrangians which yield third order EL equations; *viz.,*

$$L_{4R} := (g)^{½} G_{4R} \delta^{hi}_{rs} \varphi^r_h \varphi^s_i \qquad \text{Eq.5.13}$$

and

$$L_{5R} := (g)^{½} G_{5R} \delta^{hij}_{rst} \varphi^r_h \varphi^s_i \varphi^t_j \qquad \text{Eq.5.14}$$

where $G_{4R}$ and $G_{5R}$ are differentiable functions of $\varphi$ and $\rho$. $L_{4R}$ and $L_{5R}$ look similar to $^{BH}L_4$ and $^{BH}L_5$ given in Eqs.5.9 and 5.10, but they are different Lagrangians. In fact $L_{4R}$ and $L_{5R}$ lie inside of $^{BH}L_4$ and $^{BH}L_5$ respectively. The Lagrangians $L_{4R}$ and $^{BH}L_4$ are special cases of a class of Lagrangians investigated by Langlois and Noui in [32].

So right now we have nine second order scalar-tensor Lagrangians which yield EL equations which are of third order in the derivatives of $\varphi$ and $g_{ab}$. And, I hate to say it, but there are a few more such Lagrangians lurking in the decomposition of $L_{PD}$.

$L_{PD}$, given in Eq.5.3, can be rewritten as

$$L_{PD} = (K\varepsilon^{rstu}\varphi\varphi^p \varphi_{qt} R_p{}^q{}_{rs})_{|u} - K_\varphi \varepsilon^{rstu}\varphi\varphi_u \varphi^p \varphi_{qt} R_p{}^q{}_{rs} + \tfrac{1}{2}\varphi I - \varphi II - 2\varphi III \qquad \text{Eq.5.15}$$

where

$$I := K\varepsilon^{rstu} \varphi^p \varphi_l R_q{}^l{}_{tu} R_p{}^q{}_{rs} \qquad \text{Eq.5.16}$$

$$II := K\varepsilon^{rstu}\varphi^p{}_u \varphi_{qt} R_p{}^q{}_{rs} \qquad \text{Eq.5.17}$$

and

$$III := K_\rho \varepsilon^{rstu}\varphi^l \varphi_{lu} \varphi^p \varphi_{qt} R_p{}^q{}_{rs} . \qquad \text{Eq.5.18}$$

So Lagrangians of the $L_{PD}$ type can be written as the sum of a divergence and Lagrangians of the type I, II and III. One would hope that there exists a dimensionally dependent identity that equates I to some multiple of $L_P$, but there is no such identity, since these two Lagrangians are actually quite different. If we use Eq.3.8 to compute the fourth order terms in $E^{ab}(I)$ and $E^{ab}(II)$, we find that those terms vanish. So we can now use Eq.5.15 to deduce that there are no fourth order terms in $E^{ab}(III)$. Using Eqs.3.9 and 5.15 we can show that there are no fourth order terms in $E(I)$, $E(II)$ and $E(III)$, and that their third order terms are different. So Lagrangians I, II and III are distinct, and each yield third order scalar-tensor EL equations. Eq.5.15 allows us to conclude that the Lagrangians $L_{PD}$, I, II and III are not independent Lagrangians. Hence our analysis of $L_{PD}$ has produced two more second order scalar-tensor

Lagrangians which yield third order EL equations, which we take to be the Lagrangians I and II. And, who knows, the disformal transformations of I and II might produce a few more such Lagrangians.

The analysis which we just performed on $L_{PD}$ can now be applied to the Lagrangians $L_{GBD1}$ and $L_{GBD2}$. Fortunately it does not lead to anything new.

Thus we see that there are no shortage of second order scalar-tensor Lagrangians, which yield third order EL equations, eleven in fact so far. Things get even worse if we add one of the four different types of the second order scalar-tensor Lagrangians presented in [2], to each of the Lagrangians which yield third order EL equations.

If there is a glimmer of hope in this sea of Lagrangians, it is the fact that there are only two subclasses of Lagrangians which are conformally invariant. These Lagrangians are given in Eqs.5.2 and 5.4. Perhaps this might be useful to people doing conformal field theory. It should be noted that among the Lagrangians presented in [2] which generate second order scalar tensor field equations, there is only one that generates conformally invariant field equations; *viz.,*

$$L_{2'} := k\, (g)^{1/2}\, \rho^2$$

where *k* is an arbitrary differentiable function of φ,

I really do not know if it would be worth while to compute all possible third order scalar-tensor field equations, which are derivable from a variational principle, since there will be a plethora of them. But if you wished to try, this is how I would go about doing it.

The first thing that needs to be done is to compute all of the EL tensors corresponding to the eleven Lagrangians we have so far. By examining these tensors we should be able to determine if we actually have eleven independent Lagrangians- hopefully not. Then we need to compute all symmetric tensor densities

$$A^{ab} = A^{ab}(g_{rs}; g_{rs,t}; g_{rs,tu}; g_{rs,tuv}; \varphi; \varphi_{,r}; \varphi_{,rs}; \varphi_{,rst})$$

which are such that

$$A^{ab}{}_{|b} = \varphi^a A \qquad \text{Eq.5.17}$$

where A is a third order scalar density. If $A^{ab}$ arises from the variation of a Lagrangian of arbitrary differential order in $g_{ab}$ and $\varphi$, then it must satisfy Eq.5.17, due to Eq.2.11. Once all such $A^{ab}$ are determined, we compare it with all of the known third order EL tensors constructed above through a variation of $g_{ab}$, along with the second order scalar-tensor EL tensors presented in [2], to see if we have found anything new. If not, we are done. If there are new terms in $A^{ab}$, we shall denote those terms by $A_N{}^{ab}$. We now want a Lagrangian $L_N$ for which $E^{ab}(L_N) = A_N{}^{ab}$. Our first choice for such a Lagrangian is $g_{ab}A_N{}^{ab}$. If parts of $A_N{}^{ab}$ are trace-free, we shall denote those terms by $A_{N'a}{}^b$. To find a Lagrangian for $A_{N'}{}^{ab}$ compute $A_{N'a}{}^b{}_{|b} = \varphi^a A_{N'}$. $L' := A_{N'}$ will be our candidate for a Lagrangian with the property that $E^{ab}(L_{N'}) = A_{N'a}{}^{b}$. If $g_{ab}A^{ab}$ and $A_{N'}$ do not work, then perhaps the new terms do not come from a variational principle. Recall that the symmetric, trace-free, divergence-free Cotton tensor, does not come from the variation of a Lagrangian built from the metric tensor, even though it satisfies all the necessary conditions to do so. But I think that these two Lagrangians will yield the desired result, thereby solving the problem of constructing all possible third order scalar-tensor field equations which can be derived from a variational principle in a space of 4-dimensions.

In [7] Gleyzes, *et al.,* show that the third order terms in $^{BH}L_4$ and $^{BH}L_5$ do not lead to problems with ghosts for certain choices of $F_4$ and $F_5$. The ghost problem is also addressed by Deffayet, *et al.,* in [4] and by Langlois, and Noui in [33]. I have no idea if there will be an issue with ghosts, for the nine classes of Lagrangian I presented here.

In conclusion I must admit that I have a particular fondness for the Lagrangian φP, which essentially began our journey into the world of third order, scalar-tensor field theories. It is a remarkably simple Lagrangian, with geometrical significance. Perhaps it might be interesting to know what type of field theory develops from the addition of a constant multiple of φP to the usual Lagrangian employed to obtain the Einstein vacuum field equations with cosmological term. So we would essentially be studying the constrained variation of the usual Einstein Lagrangian subject to the constraint that P = 0. Some physicists might find such an investigation uninteresting since there would be no term involving the kinetic energy of φ in the Lagrangian. For them I suggest adding the term $k(g)^{½}ρ^2$ to the Lagrangian, where k is a constant. This additional term is conformally invariant, and hence all terms involving the scalar field in the Lagrangian would be so.

**ACKNOWLEDGEMENTS**

I wish to thank Dr. A.Guarnizo Trilleras for presenting me with a copy of his Ph.D. thesis [34]. This thesis provided me with an introduction to the efforts of those people who were trying to find third-order scalar-tensor field theories, the so-called "Beyond Horndeski Theories." Well, so long as I am alive, no one goes beyond Horndeski without me! And that explains the inception of this paper.

Professor S.Liberati informed me of the roll that **Prop. 2.1** plays in Conformal Field Theory, and for that I am very grateful.

In the course of revising this paper I have been fortunate enough to have numerous discussions with Professors C.Deffayet, D.Langlois, J.M.Ezquiaga, and Dr.

M.Zumalacárregu. These discussions have provided me with a great deal of insight into the recent developments in third-order scalar-tensor theories, and of the numerous accomplishments of these scholars in this area.

I also wish to thank Professor J.Wainwright for encouraging me to investigate the topics presented in this work.

Lastly I wish to thank my wife, Dr. Sharon Winklhofer Horndeski, for assistance in preparing this manuscript.**BIBLIOGRAPHY**

[1]   G.W.Horndeski, and D. Lovelock, "Scalar-Tensor Field Theories," Tensor **24** (1972),77-82.

[2]   G.W.Horndeski, "Second-Order Scalar-Tensor Field Equations in a Four-Dimensional Space," International J. of Theoretical Physics, **10** (1974), 363-384.

[3]   M.Zumalacárregui and J. Garcia-Bellido,"Transforming gravity: from derivative couplings to matter to second-order scalar-tensor theories beyond Horndeski," Phys.Rev.D **89** (2014), 064046, arXiv.org/abs/1308.4685, August, 2013.41

[12] G.W.Horndeski, "Dimensionally Dependent Divergences and Local Parallelizations," Tensor **21** (1973), 83-90.

[13] S.Kobayashi & K.Nomizu, "Foundations of Differential Geometry, Volume II," John Wiley & Sons, 1963.

[14] M.Ostrogradski, "Memoires sur les equations differentielles relatives au problems des isoperimetrics," Mem. Ac. St.Petersbourg **4** (1850), 385.

[15] D.Bettoni and M.Zumalacárregui, "Shaken, not stirred: kinetic mixing in scalar-tensor theories of gravity," arXiv.org/abs/1502.02666, April, 2015.

[16] J.M.Ezquiaga, J.García-Bellido, and M.Zumalacárregui, "Towards the most general scalar-tensor theories of gravity: a unified approach in the language of differential forms," Phys.Rev. D **94** (2016) 024005, arXiv.org/abs/1603.01269, March, 2016.

[17] D. Sáez-Gómez, "Scalar-Tensor Theory with Lagrange Multipliers: A way of understanding the cosmological constant problem and future singularities," arXiv.org/abs/1110.6033, January, 2012.

[18] H.A.Buchdahl, "Non-Linear Lagrangians and Cosmological Theories," Mon. Notices Roy.Astron.Soc. **150** (1970), 1.

[19] T.P.Sotiriou & V.Faraoni, "f(R) Theories of Gravity," arXiv.org/abs/0805.1726, June, 2010.
43

Geometry," Phys. Rev. D **48** (1993), 3641.

[30] D.Bettoni & S.Liberati, "Disformal Invariance of 2$^{nd}$ Order Scalar-Tensor Theories: Framing the Horndeski Action," arXiv.org/abs/1306.6724, October,2013.

[31] D.Lovelock, "Dimensionally Dependent Identities," Proc. Cambridge Philos. Soc., **68** (1970), 345-350.

[32] D.Langlois and K.Noui, "Hamiltonian analysis of higher derivative scalar-tensor theories," arXiv.org/abs/1512.06820, December, 2015.

[33] D.Langlois and K.Noui, "Degenerate Higher Order Theories beyond Horndeski: evading the Ostrogradski instability," arXiv.org/abs/1510.06930, March, 2016.

[34] A.Guarnizo Trilleras, "A Model Independent Approach to Dark Energy Cosmologies: Current and Future Constraints." Ph.D thesis, Ruprecht-Karls-Universität Heidelberg, 2015.